\documentclass[useAMS,usenatbib]{mn2e}

\usepackage{graphicx}
\usepackage{color}

\title[The orbital period of QU Carinae]{The orbital period of the \\V Sge star candidate QU Carinae
\thanks{Based on observations obtained at Observat\'orio do Pico dos Dias, LNA/MCT, Brazil.}}
\author[A. S. Oliveira et al.]{A. S. Oliveira$^{1}$\thanks{E-mail:
alexandre@univap.br (ASO)}, H. J. F. Lima$^{1}$, J. E. Steiner$^{2}$, B. W. Borges$^{3}$ and D. Cieslinski$^{4}$
\\
$^{1}$IP\&D, Universidade do Vale do Para\'iba, CEP 12244-000, S\~ao Jos\'e dos Campos, SP, Brasil\\
$^{2}$Instituto de Astronomia, Geof\'{\i}sica e Ci\^encias Atmosf\'ericas, Universidade de S\~ao Paulo, 05508-900, S\~ao Paulo, SP, Brasil\\
$^{3}$Universidade Federal de Santa Catarina, Campus Ararangu\'a, 88905-120, Ararangu\'a, SC, Brasil. \\ 
$^{4}$Divis\~ao de Astrof\'isica, Instituto Nacional de Pesquisas Espaciais,  S\~ao Jos\'e dos Campos, SP, Brasil\\
 }
\begin{document}

\date{}

\pagerange{\pageref{firstpage}--\pageref{lastpage}} \pubyear{2002}

\maketitle

\label{firstpage}

\begin{abstract}

Close Binary Supersoft X-ray Sources (CBSS) are considered strong candidates to SN~Ia progenitors, but very few CBSS are known in our Galaxy. The galactic counterparts
of the CBSS may be the V Sge stars, not detected in X-rays due to the strong absorption by the interstellar gas. Nevertheless, the number of 
members in the V Sge class is still small. In order to find new members we selected QU Car for photometric and spectroscopic observations.
The orbital period of this system was published in the literature as 10.9 h, determined from radial velocity data taken in 1979--1980, but posterior analysis
of data taken in 2006--2007 did not confirm this period. We analysed the high variability of its emission line profiles with the Temporal Variance Spectrum (TVS)
technique. Besides, we recovered the 10.9 h orbital period from the radial velocities of the \mbox{He\,{\sc ii}} 4686 {\AA} emission line and, for the first time, detected 
what may be the orbital modulation in the photometric data. This photometric modulation is present only in the lower brightness state data, when the flickering 
is attenuated. The inclusion of QU Car in the V Sge class is supported by many
features like high/low states, strong winds, nebular lines and \mbox{He\,{\sc ii}} 4686 {\AA}/H$\beta$ line ratios. However, the non-detection of the characteristic 
\mbox{O\,{\sc vi}}~3811--34 {\AA} lines in its spectrum claims against this classification. These lines, though, may be highly variable so additional spectra analysed
with the TVS technique can, possibly, solve this question.
\end{abstract}


\section{Introduction}

Despite the major importance of the Type Ia Supernovae (SN Ia), as cosmological distance indicators, to the discovery of the accelerated expansion of the Universe and the dark energy
\citep{1998AJ....116.1009R,1999ApJ...517..565P}, the unknown nature of its progenitors is still a great concern. It is a theoretical consensus that 
the progenitors are binary systems with a massive C/O white dwarf (WD) which ignites when it reaches the Chandrasekhar mass limit. However, the nature of the companion star
is under heavy debate yet. It can be a non-degenerate companion star in the 
single degenerate (SD) channel or a WD companion in the double degenerate channel. See \citet{maoz} and \citet{2012NewAR..56..122W} for 
recent reviews on the topic of SN Ia progenitors.

In the single degenerate channel, the Close Binary Supersoft Sources (CBSS) and the V Sge stars are strong candidates to SN Ia progenitors, given
their massive WD and high accretion rates. In CBSS \citep{1997ARA&A..35...69K} the WD experiences hydrostatic 
nuclear burning on its surface because of the very high ($\dot{M}\sim 10^{-7}$ M$_{\sun}~$yr$^{-1}$) accretion rate from the near main 
sequence secondary star. Firstly discovered as supersoft X-ray sources in the Magellanic Clouds, only two CBSS have been discovered in the Galaxy so far (namely QR And and MR Vel). To address the problem 
of the discrepancy between the number of CBSS discovered in other galaxies and in the Milk Way, \citet{1998PASP..110..276S} proposed that the V Sge class of stars could be the galactic
counterpart of the CBSS, not recognized as such in our Galaxy because of the absorption of the supersoft emission by the interstellar gas.

The V Sge stars are spectroscopically characterized by high-ionization emission lines of \mbox{O\,{\sc vi}} and \mbox{N\,{\sc v}} and
by the ratio of the equivalent widths of \mbox{He\,{\sc ii}} 4686 {\AA} to H$\beta$ usually greater than 2. Other characteristic features are P Cyg profiles, indicating strong wind
in the systems, and the lack or weakness of \mbox{He\,{\sc i}} lines. Their orbital periods range from 5 to 12 h and their orbital light curves are either low-amplitude sinusoidal or
high-amplitude asymmetric with primary and secondary eclipses. As in the CBSS, no spectroscopic evidence of atmospheric absorption features 
from the secondary star has been found until now in the V Sge stars, although in V Sge itself the discovery of narrow Bowen fluorescence emission features of \mbox{O\,{\sc iii}}
3132 and 3444 {\AA} characterized it as a double-lined spectroscopic binary, allowing the estimate of the mass ratio \citep{1965ApJ...141..617H}.
Four members initially composed the V Sge class: V Sge \citep{1965ApJ...141..617H}, V617 Sgr \citep{1999A&A...351.1021S,2007A&A...471L..25S}, WX Cen \citep{2004MNRAS.351..685O} and DI Cru
(HD104994, WR46), but the latter has left the V Sge class being re-classified as a quasi Wolf--Rayet (qWR) star \citep{2004PASP..116..311O}.
Attempts were made to increase the number of known
members in the V Sge class, but until now the low number of members still remains. The candidates WR 7a \citep{2003MNRAS.346..963O} and HD 45166 \citep{2005A&A...444..895S}
have also been classified as qWR stars, as in the case of DI Cru. \citet{2007MNRAS.382.1158W} noted the spectral similarity of IPHAS J025827.88+635234.9 to V Sge and to the cataclysmic variable
QU Car, however \citet{2014Ap&SS.349..361K} claims its photometric behaviour does not fit the typical behaviour of the V Sge stars and CBSS.
 Recently, \citet{2008AJ....135.1649K} (hereafter KAH08) suggested the inclusion of QU Car in the V Sge class.

QU Car, although being bright ($m_v \sim 11.4$) and known since 1968, is very poorly understood. It was discovered as an irregular variable with observational features similar to Sco X-1 by
\citet{1968ApL.....1..247S}. \citet{1969ApJ...157..709S} suggested its classification as an old nova and reported flickering as large as 0.2 mag in time-scales of minutes, but found no periodicity
in the light curve that could be associated to orbital motions. This author also reported that the spectroscopic emission lines are weakest and absorption are strongest during times of 
maximum flickering activity. \citet{1982ApJ...261..617G} (hereafter GP82) determined an orbital period of 10.9 h from the radial velocities of the \mbox{He\,{\sc ii}} 4686 {\AA}
emission line, a dominant optical spectral feature besides the \mbox{C\,{\sc iii}}/\mbox{N\,{\sc iii}} complex at 4630--4660 {\AA}. No spectral features of a secondary star were found in their
spectra, despite the long orbital period, leading to the suggestion that the light of the system is dominated by the WD and accretion disc. They suggested the classification of QU Car as 
Nova-like. While GP82 set a lower limit of 500 pc to the distance to QU Car, \citet{2003MNRAS.338..401D} estimated a possible distance of 2 kpc. If this is really the case, the luminosity would be 
$10^{37}$ erg~s$^{-1}$ and the mass accretion rate would be close to $10^{-7}$ M$_{\sun}$~yr$^{-1}$, which are typical values of the CBSS.
 
Nevertheless, a comparison made by \citet{2003MNRAS.338..401D} between the optical spectrum they obtained for QU Car and CBSS published spectra, showed that the absence of \mbox{O\,{\sc vi}}
in the former indicates that the degree of ionization in QU Car is lower than in CBSS. Based on the spectrum they also argue that the secondary may be an early-type R star. 

KAH08 obtained new optical spectra of QU Car and performed a radial velocity analysis of the \mbox{He\,{\sc ii}} 4686 {\AA} line in order to provide a modern ephemeris, 
but surprisingly they could not find, in their data set, the 10.9 h orbital period previously determined by GP82 in data taken 27 years before, using the same emission line. KAH08
propose that line profile variations due to an erratic wind may be responsible for the non-detection of the 10.9 h periodicity. In their spectra they also found signals of
the forbidden [\mbox{O\,{\sc iii}}] and [\mbox{N\,{\sc ii}}] emission lines, indicative of a nebula. This may possibly be related to the presence of a strong wind.

Based on the similarity of the spectra of QU Car to that of the V Sge star WX Cen, KAH08 proposed its inclusion on the V Sge class. Besides, they analysed
a long term AAVSO photometric time series of QU Car and discovered high and low brightness states similar to the ones presented by the V Sge stars. The same set of AAVSO
data, plus ASAS photometric monitoring, were analysed by \citet{2012MNRAS.425.1585K} (hereafter KHW12). In that data QU Car presented high states with $m_v \sim 11.5$ mag and less frequent
low states lasting for $\sim 100$ d when the magnitude was below 12 mag. KHW12 proposed to link QU Car to the V Sge class and to the Accretion Wind Evolution (AWE)
model \citep{2003ApJ...598..527H} which reproduces the high/low brightness levels as well as the off/on soft X-rays states of V Sge. 

In an observational program to search for galactic counterparts of the CBSS and new members of the V Sge class, we selected QU Car for photometric and spectroscopic studies.
In this paper we present our efforts to better understand this elusive system. In section 2 we present our data and in section 3 we discuss the extensive series of spectra,
both in terms of line profile and radial velocity variabilities, and also discuss the photometric data. Our conclusions are presented in section~4. 

\section{Observations and data reduction}

QU Car was observed with all three telescopes at the Observat\'orio Pico dos Dias (OPD -- LNA/MCTI) located in southeast Brazil.
Photometric observations were carried out on ten nights during 2009, 2011 and 2012 (Table~\ref{jophoto}) at both 0.6-m Zeiss and Boller \& Chivens telescopes.
We employed three distinct thin, back-illuminated detectors: the E2V CCD47-20 (CCD S800), the SITe SI003AB (CCD 101) and the E2V CCD42-40 (CCD IkonL).
Time series of images were obtained through the Johnson V filter, with individual exposure times of  30 seconds. The timings were provided by a GPS 
receiver. Bias and dome flatfield exposures were used for correction of the detector read-out noise and sensitivity using standard 
\textsc{iraf}\footnote {\textsc{iraf} is distributed by the National Optical Astronomy Observatories, which are operated by the Association of Universities for 
Research in Astronomy, Inc., under cooperative agreement with the National Science Foundation.} tasks.
The differential aperture photometry was performed with the DAOPHOT II package using apertures and background annulus defined by the instantaneous 
PSF measured at each image.

\begin{table}
 \centering
  \caption{Journal of photometric observations of QU~Car.\label{jophoto}}
  \begin{tabular}{@{}lccc@{}}
  \hline
   Date   & Length of        &  Telescope & CCD \\
   (UT)   & observation (h)  &            &     \\
  \hline
2009 Jun 06	& 2.8   & 0.6 m B\&C	& S800 \\
2011 Apr 13	& 7.7   & 0.6 m Zeiss	& 101  \\
2011 Apr 22	& 6.3   & 0.6 m Zeiss	& 101  \\
2011 Apr 23	& 3.5   & 0.6 m Zeiss	& 101  \\
2011 Apr 24	& 3.7   & 0.6 m Zeiss	& 101  \\
2011 May 11	& 1.7   & 0.6 m Zeiss	& IkonL\\
2011 Jun 21	& 7.0   & 0.6 m Zeiss	& IkonL\\
2012 Mar 14	& 8.1   & 0.6 m Zeiss	& 101  \\
2012 Mar 25	& 9.3   & 0.6 m Zeiss	& 101  \\
2012 Mar 26	& 6.0   & 0.6 m Zeiss	& 101  \\
\hline
\end{tabular}
\end{table}

The spectroscopic data were obtained with the 1.6-m Perkin--Elmer and the 0.6-m Boller \& Chivens telescopes at OPD in 12 nights during 2004, 2008, 2010 and 2012.
Table~\ref{jospec} presents a journal of the spectroscopic observations. Two thin, back-illuminated Marconi detectors (CCD 098 and CCD 105) were used with the 
Coud\'e or Cassegrain spectrographs. Bias and flatfield corrections were applied as usual. The width of the slit was adjusted to the conditions of the seeing. 
We took exposures of calibration lamps after every third exposure of the star, in order to determine accurate wavelength calibration solutions. 
The image reductions, spectra extractions and wavelength calibrations were executed with \textsc{iraf} standard routines.  

\begin{table*}
 \centering
 \begin{minipage}{140mm}
  \caption{Journal of spectroscopic observations of QU~Car.\label{jospec}}
  \begin{tabular}{@{}ccccccccc@{}}
  \hline
   Date     &  Telescope & Spectrograph & Grating       & CCD & Exp. Time & Number   & Resol.   & Spec. range \\
   (UT)     &            &              & (l mm$^{-1}$) &     & (s)       & of exps. & ({\AA})  &  ({\AA})    \\
  \hline
2004 Feb 17 & 1.6 m P-E   & Coud\'e        & 600        & 098 & 600       & 5        &   0.7    & 4040--5170 \\
2004 Mar 01 & 1.6 m P-E   & Coud\'e        & 600        & 098 & 1200      & 8        &   0.7    & 4040--5170 \\
2004 Mar 11 & 0.6 m B\&C  & Cass.          & 900        & 105 & 900       & 9        &   2      & 3800--5230 \\
2008 Apr 02 & 1.6 m P-E   & Coud\'e        & 600        & 105 & 900       & 1        &   0.6    & 4500--5000 \\
2010 Feb 13 & 1.6 m P-E   & Coud\'e        & 600        & 098 & 600       & 26       &   0.6    & 4205--5340 \\
2010 Feb 14 & 1.6 m P-E   & Coud\'e        & 600        & 098 & 600       & 27       &   0.6    & 4205--5340 \\
2010 Feb 16 & 1.6 m P-E   & Coud\'e        & 600        & 098 & 600       & 17       &   0.6    & 4205--5340 \\
2012 Mar 01 & 1.6 m P-E   & Coud\'e        & 600        & 098 & 1200      & 4        &   0.7    & 4000--5140 \\
2012 Mar 02 & 1.6 m P-E   & Coud\'e        & 600        & 098 & 1200      & 8        &   0.7    & 4000--5140 \\
2012 Mar 03 & 1.6 m P-E   & Coud\'e        & 600        & 098 & 1200      & 1        &   0.7    & 4000--5140 \\
2012 Mar 04 & 1.6 m P-E   & Coud\'e        & 600        & 098 & 1200      & 6        &   0.7    & 4000--5140 \\
2012 Mar 05 & 1.6 m P-E   & Coud\'e        & 600        & 098 & 1200      & 21       &   0.7    & 4000--5140 \\
\hline
\end{tabular}
\end{minipage}
\end{table*}

\section[]{Data analysis and results}


\subsection{Variability in the spectral features}

\begin{figure}
\resizebox{\hsize}{!}{\includegraphics[clip]{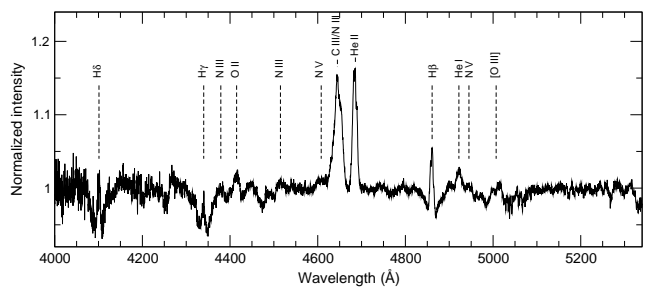}}
\vspace*{10pt}
\caption{Average normalized spectrum of QU Car constructed with 2004, 2010 and 2012 OPD data. The positions of identified features are indicated. \label{averspec}}
\end{figure}

\begin{figure}
\vspace*{10pt}
\centerline{\includegraphics[width=84mm]{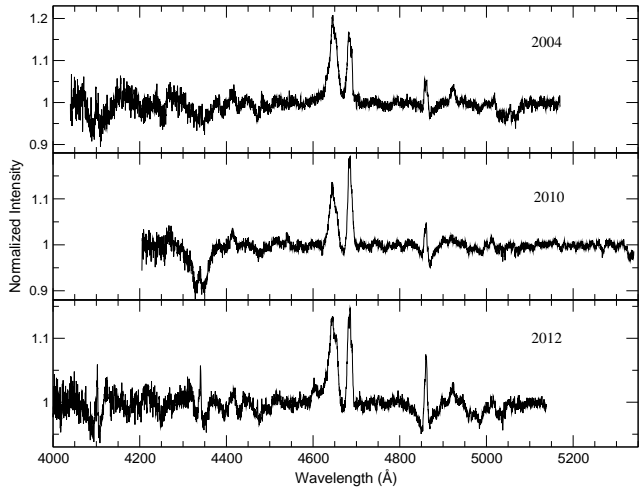}}
\caption{ Average normalized spectra of QU Car in 2004, 2010 and 2012. \label{3spec}}
\end{figure}

The average spectrum of QU Car constructed with data obtained in 2004, 2010 and 2012, normalized to the continuum, is presented in Fig.~\ref{averspec}. It is dominated by the emission 
features of \mbox{He\,{\sc ii}} 4686 {\AA}
and the \mbox{C\,{\sc iii}}/\mbox{N\,{\sc iii}} complex. The Balmer emission lines are superposed on broad absorption features. Emission lines of \mbox{N\,{\sc v}} 4603, 4619 
and 4945 {\AA}, \mbox{N\,{\sc iii}} 4379 and 4515 {\AA},  \mbox{O\,{\sc ii}} 4415 {\AA},  \mbox{He\,{\sc i}} 4922 {\AA} and [\mbox{O\,{\sc iii}}] 
5007 {\AA} are also present. The features in the spectra are highly variable, both in terms of intensities and profiles. Fig.~\ref{3spec} shows the average spectra for 2004, 2010 and 2012, and Table~\ref{lines} 
lists the equivalent widths and FWHM of the emission lines in the average spectra of these years.   

The broad Balmer absorption troughs are variable and were also observed by GP82, \citet{2003MNRAS.338..401D} and KAH08. A possible explanation for the origin of these troughs could be the 
spectral features of the white dwarf atmosphere, although it would imply that the emission of the primary has a substantial contribution in the optical, in contrast to other evidences 
of a high accretion rate occurring in QU Car. A more convincing explanation is that the Balmer troughs are formed in the optically thick accretion disc. Such absorptions are observed 
in the spectra of dwarf novae during eruption and in UX UMa nova-likes (so called thick-disc CVs) as RZ Gru \citep{Warner}. During the evolution to the maximum of the dwarf nova eruption
there is a transition from emission-line to absorption-line spectrum, as well described for SS Cyg \citep{1984ApJ...286..747H}. An important point to note is that in such cases
the Balmer decrement is much steeper in the emission than in the absorption lines, resulting in H$\alpha$ in emission while higher series members present progressively 
stronger absorptions \citep{Warner}. This seems to be the case observed in the spectrum of QU Car presented, for instance, in fig. 1 of KAH08.

The ratio of the equivalent widths of the \mbox{He\,{\sc ii}} 4686 {\AA} to H$\beta$ emission lines, which is typically greater than 2 in CBSS and V Sge stars, varied from 
$EW_{\mbox{He\,{\sc ii}}}/EW_{H\beta}=2.4$ in our 2004 data to 4 in 2010 and back to 2 in 2012, while GP82 measured $EW_{\mbox{He\,{\sc ii}}}/EW_{H\beta}=2$ in 1979 
and KAH08 obtained a value 
always lower than 1 in 2006 and 2007 data. 
These ratios, however, are uncertain and should be looked with caution, since the presence of the variable Balmer absorption troughs in all observed spectra
of QU Car affects the measurements of the EW of the H$\beta$ emission. We, like KAH08, measured the EW of the Balmer lines only in their emission cores.
The ratio of the EW of the \mbox{He\,{\sc ii}} to the Bowen \mbox{C\,{\sc iii}}/\mbox{N\,{\sc iii}} complex, on the other hand,
changed from 0.5 to 0.9 and to 0.6 between 2004, 2010 and 2012.

In the 2012 average spectrum the \mbox{N\,{\sc v}} 4603 and 4619 {\AA} emission lines are present, while \mbox{N\,{\sc v}} 4945 {\AA} is visible in the average spectra of 2010 and 2012. These
high-ionization lines, together with the strong \mbox{He\,{\sc ii}} 4686 {\AA} line, are defining features of the V Sge stars. But, differently from what happens 
in the V Sge stars, in QU Car the \mbox{He\,{\sc i}} 4922 {\AA} emission line is present, as can be seen in its 2004 and 2012 average spectra. \citet{1998PASP..110..276S} analysed the presence 
of \mbox{He\,{\sc ii}} and the absence of \mbox{He\,{\sc i}} in the V Sge stars and suggested that these lines are formed by photoionization in a matter limited region, in contrast
with the radiation limited case often found in Cataclysmic Variables. In this context, it is interesting to note the possible anti-correlation between the intensities of the \mbox{He\,{\sc ii}} 4686 {\AA}
and \mbox{He\,{\sc i}} 4922 {\AA} lines in our QU Car spectra: while in 2004 \mbox{He\,{\sc ii}} is less intense than the Bowen complex, the \mbox{He\,{\sc i}} line is quite prominent 
($EW_{\mbox{He\,{\sc i}}} = -0.5$). This situation is inverted in 2010, when \mbox{He\,{\sc ii}} is more intense than the Bowen complex and the \mbox{He\,{\sc i}} line is marginally detected
($EW_{\mbox{He\,{\sc i}}} = -0.1$). In 2012 an intermediate situation occurs.

In order to examine the variability of the emission line profiles in more detail, we performed the temporal variance spectrum (TVS) analysis on our spectroscopic data. In this procedure, the temporal
variance is calculated for each wavelength pixel, from the residuals of each continuum normalized spectrum to the average spectrum. In our implementation, the temporal variance spectrum is
the square root of the variance as a function of wavelength.
A characteristic indicator of the variability of each spectral feature is the ratio between its variance and its intensity, $\sigma/I$. 
The TVS can be a useful method to distinguish between different kinds of features present in a line spectrum. Telluric lines, for instance, can present very high values for $\sigma/I$, while 
interstellar lines should not appear in the TVS. Also, if a line has no intrinsic profile variation but has radial velocity displacement only, its TVS should present an unambiguous double peak
profile with $\sigma/I \sim K/FWHM$.
 For further details on the TVS method see \citet{1996ApJS..103..475F}. 

\begin{figure}
\vspace*{10pt}
\centerline{\includegraphics[width=1.1\columnwidth]{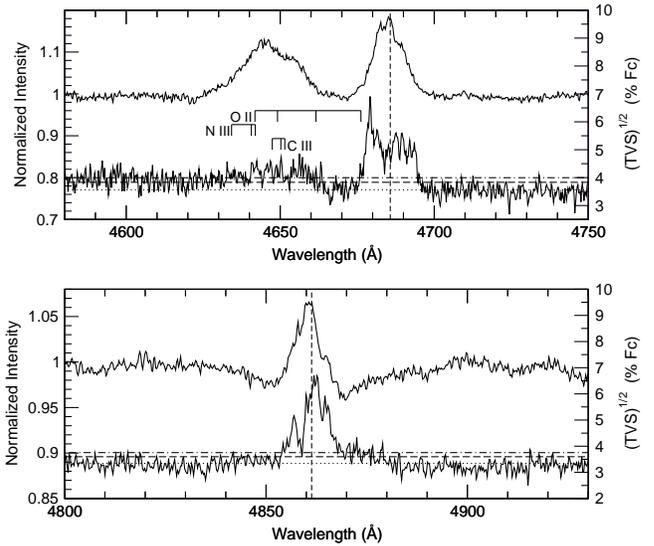}}
\caption{Average intensity spectrum and TVS for the \mbox{C\,{\sc iii}}/\mbox{N\,{\sc iii}} complex and \mbox{He\,{\sc ii}} 4686 {\AA} (upper panel) and  H$\beta$ (bottom panel).
The TVS ordinates are represented in the right axis and give the amplitudes as percentage of the 
normalized continuum. The TVS statistical threshold significance of p=1, 5 and 30 per cent are represented by dot-dashed, dashed and dotted horizontal lines, respectively. The vertical dashed lines 
mark the position of the rest velocity for the relevant line presented in each panel. Rest positions of the lines of \mbox{C\,{\sc iii}}, \mbox{N\,{\sc iii}} and 
\mbox{O\,{\sc ii}} are indicated by the bars.  \label{tvs}}
\end{figure}

We constructed the TVS for the strongest spectral lines, such
as the Bowen complex,  \mbox{He\,{\sc ii}} 4686 {\AA} and H$\beta$, using our 110 spectra obtained in 2010 and 2012 (the lower number of spectra obtained in 2004 precludes this analysis
for that year).
Fig.~\ref{tvs} shows the observed average intensity spectrum of QU Car and the calculated TVS. The most striking feature is the difference between the TVS profiles of the Bowen complex
and of the \mbox{He\,{\sc ii}} line. While the TVS of \mbox{He\,{\sc ii}} presents a rich and intense profile, the Bowen complex show marginally significant features above the 1 per cent
statistical level, representing a lower variability in this complex. In Fig.~\ref{tvs} we also show the rest positions of the strongest lines (\mbox{C\,{\sc iii}}, \mbox{N\,{\sc iii}} and 
\mbox{O\,{\sc ii}}) of the Bowen blend \citep{1975ApJ...198..641M}, which coincides with some features in the respective TVS.

The \mbox{He\,{\sc ii}} intensity emission line has an asymmetric compound profile with a main peak at 4685 {\AA} ($\Delta v = -30$ km s$^{-1}$) and some smaller side emissions. 
The TVS of \mbox{He\,{\sc ii}} is much more asymmetric, with a central dip centred at the velocity of $-120$ km s$^{-1}$. It shows strong variability at $\Delta v = -420$ km s$^{-1}$,
with $\sigma/I=6$ per cent, that seems to be associated to a blue emission component observed in the intensity spectrum. The red wing of \mbox{He\,{\sc ii}} also presents relevant variability in
the TVS, with $\sigma/I=5$ per cent, which may be associated to a red emission component at $\Delta v = +240$ km s$^{-1}$. These possible associations are clearer in the TVS
constructed with separate 2010 and 2012 data, not shown in this paper. The core of the line, on the other hand, shows lower variability, with $\sigma/I=4$ per cent.
Another possible interpretation to the behaviour in the blue wing of \mbox{He\,{\sc ii}} is a variable P Cyg profile causing maximum variance in the blue.

The H$\beta$ emission line presents a situation similar to \mbox{He\,{\sc ii}}. In the blue wing of this emission line the TVS shows a peak displaced at $\Delta v = -270$ km s$^{-1}$ with 
$\sigma/I=5$ per cent, and in this same velocity the intensity spectrum presents a clear peak. In the red wing, the TVS has a peak at $\Delta v = +220$ km s$^{-1}$ with $\sigma/I=6$ per cent, again 
at the same velocity of a emission component in the intensity spectrum. But, differently from \mbox{He\,{\sc ii}}, the central core of H$\beta$ emission exhibit high variability with 
$\sigma/I=6$ per cent. An interesting fact is that the peak of the line is displaced to $\Delta v = -40$ km s$^{-1}$ in the intensity spectrum while the peak in the TVS is displaced to
 $\Delta v = +60$ km s$^{-1}$. We do not have interpretation for these velocities.
No statistically significant variability is evident in the blue wing of the broad absorption feature of H$\beta$, while in the red wing the variability is only marginally significant at the 1 per cent level.

The [\mbox{O\,{\sc iii}}] 5007 {\AA} forbidden line, observed before in QU Car by KAH08 and KHW12, is not accompanied by 
the weaker component of the doublet at 4959~{\AA}.
In spectra obtained in 2006 and 2007, KAH08 observed this feature split in two components with velocities of $-500$ and $+370$ km s$^{-1}$, suggesting their formation in the front and 
back sides of an expanding shell. In 2010 and 2011, KHW12 observed only one weak velocity component of [\mbox{O\,{\sc iii}}] 5007 {\AA}. Our data clearly show both components in 
the average spectra taken in 2004, 2010 and 2012. Table~\ref{OIIIvel} presents the velocities of these components measured in the average 
spectra of each year. The TVS in the spectral region of the [\mbox{O\,{\sc iii}}] 5007 {\AA} line does not exhibit statistically significant variations.

\begin{table*}
\centering
\caption{\label{lines}Equivalent widths (EW) and FWHM of spectral lines of QU Car in the average spectra of 2004, 2010 and 2012.
The EW of the Balmer lines are measured only in the emission cores. For [\mbox{O\,{\sc iii}}] 5007 {\AA} we present the measurements of the 
blue and red components as well as the full emission line (see text for details).}
\begin{tabular}{@{}lcccccc@{}}
\hline
Species   & EW in          & FWHM in      & EW in                & FWHM in            & EW in        & FWHM in      \\
          &  2004 ({\AA})  & 2004 (km s$^{-1}$) & 2010 ({\AA})   & 2010 (km s$^{-1}$) &  2012 ({\AA})& 2012 (km s$^{-1}$) \\
\hline
H$\delta$                                                 & $-0.7$ & 700  & ...  & ...    & $-0.5$ & 590  \\
H$\gamma$                                                 & $-0.2$ & 430  & $-0.3$ & 570  & $-0.4$ & 400  \\
\mbox{N\,{\sc iii}} 4379 {\AA}                            & $-0.7$ & 1330 & $-0.3$ & 1030 & $-0.8$ & 2350 \\
\mbox{O\,{\sc ii}}  4415 {\AA}                            & $-0.7$ & 1360 & $-0.4$ & 980  & $-0.5$ & 1160 \\
\mbox{C\,{\sc iii}}/\mbox{N\,{\sc iii}}/\mbox{O\,{\sc ii}} 4642 {\AA}& $-4.1$ & 1420 & $-2.8$ & 1350 & $-3.0$ & 1400 \\
\mbox{He\,{\sc ii}} 4686 {\AA}                            & $-1.9$ & 770  & $-2.4$ & 740  & $-1.8$ & 720 \\
H$\beta$                                                  & $-0.8$ & 600  & $-0.6$ & 480  & $-0.9$ & 460  \\
\mbox{He\,{\sc i}} 4922 {\AA}                             & $-0.5$ & 820  & $-0.1$ & 380  & $-0.3$ & 730 \\
$[\mbox{O\,{\sc iii}}]$ 5007 {\AA} full                   & $-0.9$ & 2120 & $-0.5$ & 1640 & $-0.9$ & 1960 \\
$[\mbox{O\,{\sc iii}}]$ 5007 {\AA} blue                   & $-0.2$ & 780  & $-0.1$ & 1020 & $-0.2$ & 930 \\
$[\mbox{O\,{\sc iii}}]$ 5007 {\AA} red                    & $-0.5$ & 1050 & $-0.2$ & 610  & $-0.3$ & 760 \\
\hline
\end{tabular}
\end{table*}

\begin{table}
\centering
\caption{\label{OIIIvel} Velocities of the blue and red components of the $[\mbox{O\,{\sc iii}}]$ 5007 {\AA} emission line in the average spectra of 2004, 2010 and 2012.
}
\begin{tabular}{@{}ccc@{}}
\hline
Year      &  v$_-$           &  v$_+$                \\
          &  (km s$^{-1}$)   & (km s$^{-1}$)    \\
\hline
2004      & $-370$ & $+670$    \\
2010      & $-490$ & $+340$    \\
2012      & $-560$ & $+490$    \\
\hline
\end{tabular}
\end{table}

\subsection{The spectroscopic orbital period}

\begin{figure}
\vspace*{10pt}
\centerline{\includegraphics[width=84mm]{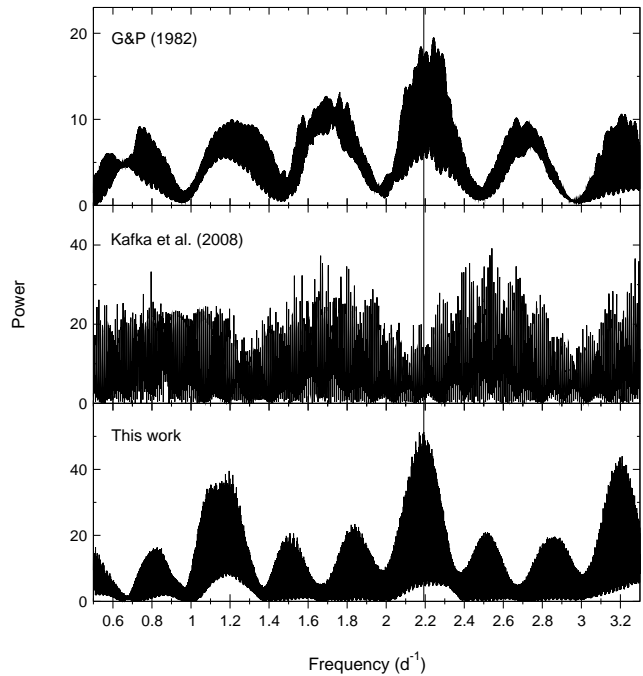}}
\caption{Lomb--Scargle periodogram of the \mbox{He\,{\sc ii}} 4686 {\AA} radial velocities from \citet{1982ApJ...261..617G} (top), \citet{2008AJ....135.1649K} (middle) and from our 2010 and 2012 
spectra (bottom). The vertical line indicates the period of 10.94 h (2.193 d$^{-1}$), which is the highest peak in our power spectrum.
 \label{lombspec}}
\end{figure}

We measured radial velocities of the \mbox{He\,{\sc ii}} 4686 {\AA} line by fitting a gaussian function to the peaks of the line profiles, and used the values obtained to search for periodicities.
Fig.~\ref{lombspec} shows the Lomb--Scargle \citep{1982ApJ...263..835S} periodogram of our 2010 and 2012 radial velocity data, which correspond to the most homogeneous set of spectra, as well as the 
periodograms we constructed from the GP82 and KAH08 \mbox{He\,{\sc ii}} 4686 {\AA} radial velocities.

\begin{figure}
\vspace*{10pt}
\resizebox{\hsize}{!}{\includegraphics[clip]{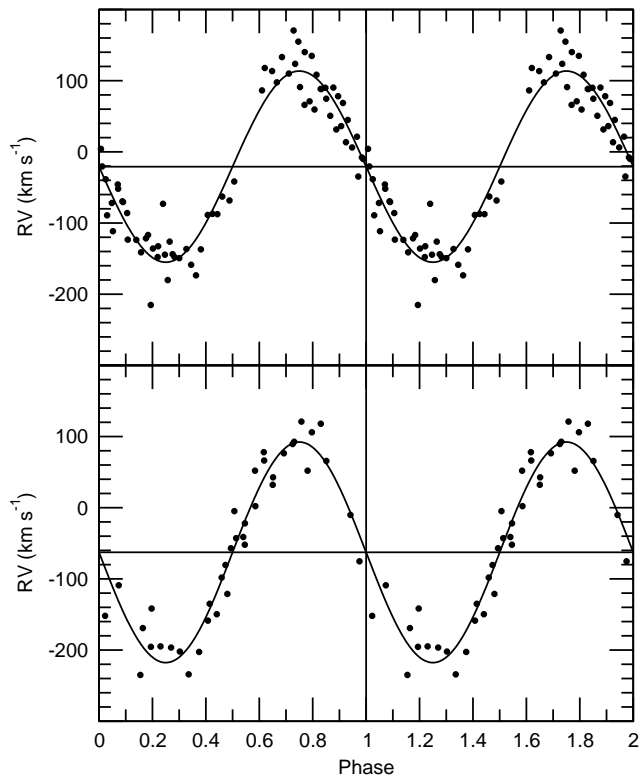}}
\caption{\mbox{He\,{\sc ii}} 4686 {\AA} radial velocity curve from 2010 (upper panel) and 2012 (lower panel) data, folded with the period and epoch given in the 
respective ephemeris. The solid curve is the sinusoidal fit to the data 
and the horizontal line represents the systemic velocity $\gamma$.
 \label{vrfase}}
\end{figure}

The individual periodograms of the three sets of data taken in 2004, 2010 and 2012 at OPD are very similar to the periodogram of the 2010/2012 combined data set shown in Fig.~\ref{lombspec}.
This result shows that the 10.9 h orbital period determined by GP82 from 1979 and 1980 data, which was later absent in radial velocity data of KAH08 taken in 2006 and 2007, 
reappears as P = 10.94 h in our spectra obtained in 2010 and 2012, being also present in our 2004 data. The spectroscopic ephemerides associated to the \mbox{He\,{\sc ii}} 4686 {\AA} emission line
from 2010 and 2012 data are shown in Table~\ref{ephem}. The zero phase is defined as the crossing from positive to negative values of the radial velocity, when compared
to the systemic velocity $\gamma$ of each set, P is the orbital period from that set and $K_1$ is the semi-amplitude of the sinusoidal fit. In Fig.~\ref{vrfase} we present the radial 
velocity curves of 2010 and 2012 data, folded with the period and epoch of each respective ephemeris.

\begin{figure}
\vspace*{10pt}
\resizebox{\hsize}{!}{\includegraphics[clip]{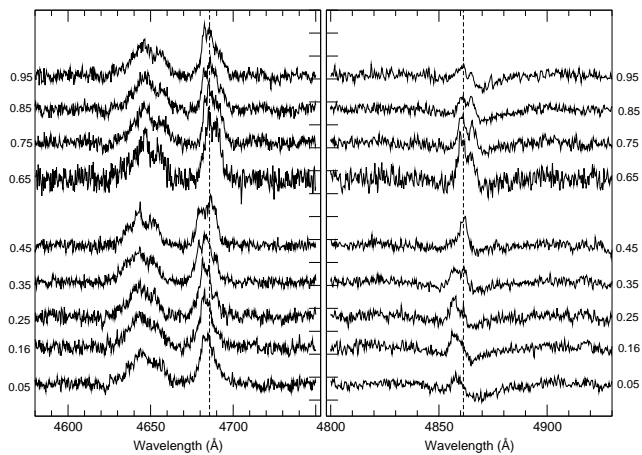}}
\caption{Average continuum subtracted 2010 spectra of the \mbox{C\,{\sc iii}}/\mbox{N\,{\sc iii}} complex and \mbox{He\,{\sc ii}} 4686 {\AA} (left panel) and  H$\beta$
 (right panel) in nine phase bins. The effective phase of each bin is indicated in the ordinate axes. The vertical dashed lines mark the position of the rest velocity of 
\mbox{He\,{\sc ii}} 4686 {\AA} and  H$\beta$.
 \label{trailedspec2010}}
\end{figure}

\begin{figure}
\vspace*{10pt}
\resizebox{\hsize}{!}{\includegraphics[clip]{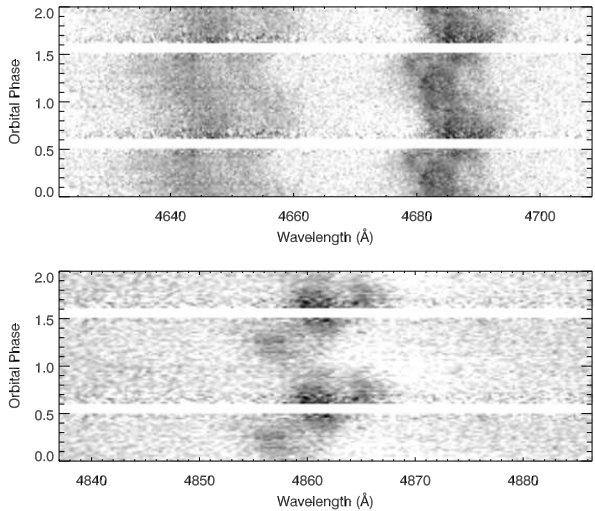}}
\caption{Trailed spectrograms of the \mbox{C\,{\sc iii}}/\mbox{N\,{\sc iii}} complex and \mbox{He\,{\sc ii}} 4686 {\AA} (upper panel) and  H$\beta$
 (lower panel) line profiles of 2010 data, binned into 0.02 phase intervals. The orbital cycle is repeated for clarity.
 \label{greenst}}
\end{figure}

Fig.~\ref{trailedspec2010} shows the binned average spectra of the \mbox{C\,{\sc iii}}/\mbox{N\,{\sc iii}} complex,
 \mbox{He\,{\sc ii}} 4686 {\AA} and  H$\beta$ lines, obtained in 2010 (our largest data set), phased with the orbital period and ${T_0}$ from the 2010 ephemeris, 
each spectrum being the average of five to ten spectra depending on the bin, while Fig.~\ref{greenst} presents the trailed spectrograms of the same data set. 
The \mbox{C\,{\sc iii}}/\mbox{N\,{\sc iii}} complex, \mbox{He\,{\sc ii}} 4686 {\AA} and  H$\beta$ lines are clearly in phase.
When comparing the structures in the line profiles of \mbox{He\,{\sc ii}} 4686 {\AA} and  H$\beta$ at the same orbital phase bin, one can see that these structures are quite similar
(see, for instance, the profiles at phase 0.65), perhaps indicating that these features are produced in the same location.
The profiles of these lines are strongly variable, specially H$\beta$ which additionally experiences the presence of the variable absorption trough. When exploring the individual spectra
we could see that, in some occasions, H$\beta$ almost completely
disappears in the noise, usually during phases 0.8 to 0.3. However, due to the low S/N ratio in the individual spectra we could not ensure whether this phenomenon is associated to the 
orbital phase or to a non-orbital source of variability, although the disappearance of H$\beta$ for about half an orbital cycle was also reported by GP82.

\begin{table*}
\caption{\label{ephem} Radial velocity parameters of the \mbox{He\,{\sc ii}} 4686 {\AA} emission line from 2010 and 2012 data. }
\begin{center}
\begin{tabular}{l c c c c}
\hline
Year &  ${T_0}$ & P      & K$_1$          & $\gamma$ \\
     &  (HJD)   & (d)  & (km s$^{-1}$)  & (km s$^{-1}$) \\
\hline
2010       & 2~455~241.785 $\pm$ 0.027 & 0.450 $\pm$ 0.009 & 134 $\pm$ 28 & $-21$ \\
2012       & 2~455~992.920 $\pm$ 0.036 & 0.456 $\pm$ 0.010 & 155 $\pm$ 30 & $-63$ \\
\hline
\end{tabular}
\end{center}
\end{table*}

\subsection{The photometric orbital period}

After seeing the recovery of the 10.9 h orbital period in our spectroscopic data, we searched for periodic modulations in the photometric data.
In order to better compare our photometric data to the high/low brightness levels published by KAH08 and KHW12, we converted our differential 
magnitudes to V magnitudes. For that we used as reference one of our differential comparison stars, C1, which is registered in the Tycho-2 catalogue \citep{2000A&A...355L..27H} as TYC 9212-2118-1, 
with $m_v = 11.089 (\pm 0.071)$.

\begin{figure}
\vspace*{10pt}
\centerline{\includegraphics[width=84mm]{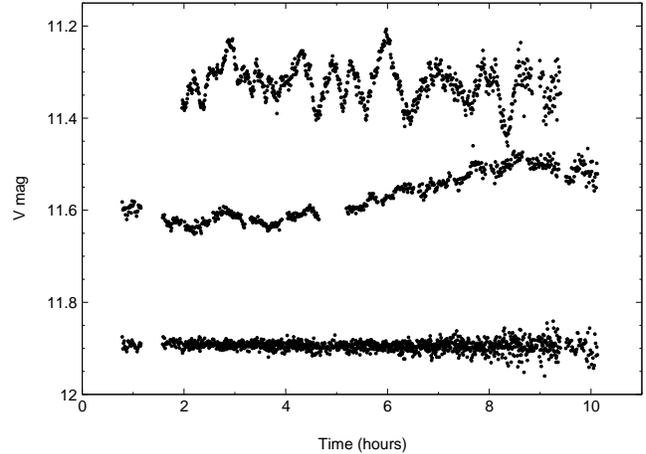}}
\caption{ Differential light curve of QU Car obtained in 2011 June 21, which is dominated by flickering (upper curve), light curve obtained in 2012 March 25, in which flickering is 
attenuated (middle curve) and the superposed light curves of the comparison star C1
for both dates (bottom curve). The vertical displacement of 0.3 mag between the light curves of QU Car is real and shows the variation in the brightness level. The 
magnitudes of the C1 are shifted by 0.37 mag and the runs are arbitrarily offset in time for 
display purposes. The gaps in the light curves were caused by clouds.
 \label{fotaltobaixo}}
\end{figure}

\begin{figure}
\vspace*{10pt}
\resizebox{\hsize}{!}{\includegraphics[clip]{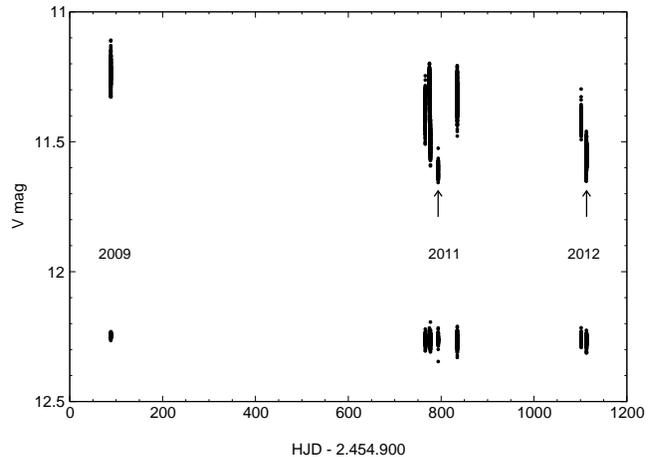}}
\caption{ Photometric data of QU Car taken at OPD in 2009, 2011 and 2012 (upper points) and of the comparison star (bottom points). The stability of the comparison star brightness shows that 
the variation of 0.5 mag seen in QU Car data, in time-scales of tens of days, is real. 
The arrows indicate the nights of 2011 May 11 and 2012 March 25 and 26 (see text for details).
 \label{fotlevels}}
\end{figure}

The light curves of QU Car show many kinds of variability, with time-scales ranging from minutes to months. The most prominent one is flickering, with time-scales of
tens of minutes and amplitudes of 0.2 mag as can be seen, for example, in the light curve of 2011 June 21 (upper light curve in Fig.~\ref{fotaltobaixo}). Superposed on this flickering there are
slow (hours) and smooth non-periodic fluctuations of tenths of magnitude. QU Car also presents different brightness levels with amplitude of about 0.5 mag and time-scales of
tens of days in our set of data (Fig.~\ref{fotlevels}), which were also reported by KAH08 and KHW12.
It is important to say that QU Car was never caught below visual magnitude 11.7 in our set of data, differently from observations described in KHW12, where it 
occasionally dropped below 12 mag.
There seems to exist in our data a correlation between the 
brightness level and the flickering activity of QU Car, which was also registered 
by \citet{1986PASP...98.1336K}. Fig.~\ref{fotaltobaixo} shows a comparison of the light curves obtained on 2011 June 21, when it was at the bright state, and on 2012 March 25
when it was at the lower state. The lower brightness state, 
which is about 0.3 mag fainter in this comparison, presents flickering with amplitude ten times smaller than the flickering observed in the higher state. In other occasions when the star was
in this lower brightness state (2011 May 11 and 2012 March 26) the same flickering attenuation was observed. 

\begin{figure}
\vspace*{10pt}
\resizebox{\hsize}{!}{\includegraphics[clip]{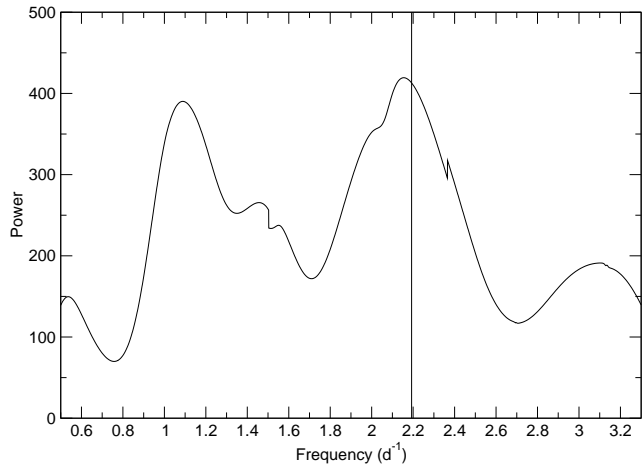}}
\caption{Lomb--Scargle periodogram of the photometric data taken in 2012 March 25 and 26. The vertical line indicates the orbital period of 10.94 h (2.193 d$^{-1}$) obtained from the power 
spectrum of the \mbox{He\,{\sc ii}} 4686 {\AA} spectral line.
 \label{lombfot}}
\end{figure}

\begin{figure}
\vspace*{10pt}
\resizebox{\hsize}{!}{\includegraphics[clip]{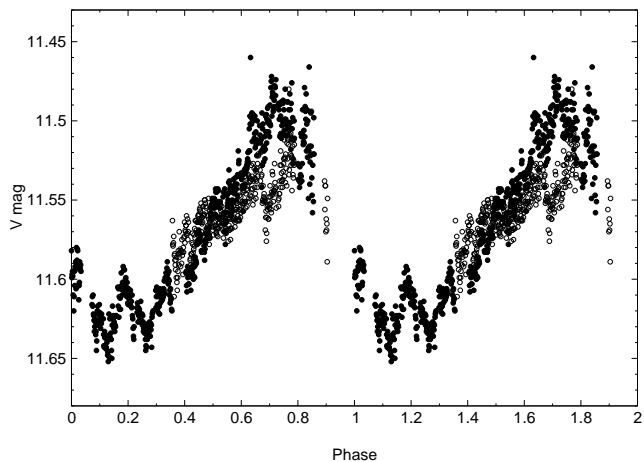}}
\caption{Light curve for the nights 2012 March 25 (filled circles) and 2012 March 26 (open circles), folded with the orbital period of 10.94 h. The length of the light curves are 9.3 h and 
6.0 h, respectively.
 \label{lcurve}}
\end{figure}

We applied the Lomb--Scargle method to search for periodicities in the photometric data. The periodograms constructed with the higher brightness level data do not show any relevant peak. 
On the other hand, despite the short length of the light curves, the periodogram of the lower state photometric data presents a period of $\sim11.1$ h
(Fig.~\ref{lombfot}) which is consistent with the 10.9 h period obtained from the \mbox{He\,{\sc ii}} 4686 {\AA} line radial velocities. In Fig.~\ref{lcurve} we show the light curves of 
2012 March 25 and 26 phased on the 10.94 h period. The amplitude of the modulation is about 0.15 mag. This appears to be the first photometric detection of an orbital modulation in
QU Car and, if so, it occurred when the amplitude of the flickering had reduced from its typical value of 0.2 mag in the higher brightness state to $\sim0.02$ mag in the lower state, unveiling 
the 0.15 mag orbital modulation. Photometric monitoring on QU Car were carried out by \citet{1969ApJ...157..709S}, GP82 and \citet{1986PASP...98.1336K} but no orbital modulation was detected.

\section{Conclusions}

An important result of this work is the recovery of the 10.9 h orbital period in the spectroscopic data of QU Car, but even more important may be the discovery of the long sought 
orbital modulation in the photometric observations. This orbital modulation has an amplitude of 0.15 mag and could be buried in the usual 0.2 mag flickering. The attenuation
of the flickering when the system is in the lower brightness state was already noted by \citet{1986PASP...98.1336K} and is also observed in our data, and it seems that this attenuation
is needed to unveil the photometric orbital modulation. It is worth to say that, in our photometry, QU Car was not seen in its lowest registered brightness level.

KAH08 suspected that the non-detection of the orbital modulation in their spectroscopic data could be related to the presence of a wind 
distorting the spectral line profiles. In addition to the line profile variations -- LPV -- the wind also manifests itself in the form of observed nebular lines and in the P Cyg profiles,
and is a key ingredient of the Accretion Wind Evolution model for CBSS and V Sge stars. As a tool to investigate the LPV we applied the TVS method to our spectra, which showed 
that the \mbox{He\,{\sc ii}} 4686 {\AA} line, although used to map the radial velocities of the system, is heavily affected by the profile variations in a complex way.

As suggested by \citet{2003MNRAS.338..401D} and KAH08 QU Car has many features that link it to the CBSS and V Sge classes, like the strong wind, nebular lines,
high and low states, high accretion rate, luminosity and \mbox{He\,{\sc ii}}/H$\beta$ line ratio. However one of the defining characteristics of the CBSS/V Sge stars are the 
\mbox{O\,{\sc vi}} 3811--34 {\AA} emission lines, which are absent in the spectrum of QU Car \citep{2003MNRAS.338..401D}. The lack of detection of \mbox{O\,{\sc vi}} is consistent 
with a lower degree of ionization in QU Car 
when compared to other CBSS/V Sge systems, but could be also due to variability in this line, like the one observed in the spectra of WX Cen \citep{2004MNRAS.351..685O}.
 The TVS method has proven to be a useful tool to enhance the signal of
strongly variable \mbox{O\,{\sc vi}} lines in noise dominated spectra of other V Sge candidates, as WR 7a \citep{2003MNRAS.346..963O} and WX Cen \citep{2004MNRAS.351..685O}.
Unfortunately our QU Car spectra do not cover the \mbox{O\,{\sc vi}} line region (except for 2004 March 11, but then the noise dominated that region), so new bluer spectra 
combined with the TVS analysis are needed to search for this important
feature. 

In order to understand a system so variable as QU Car, extensive observations with several techniques are mandatory. Long term photometric observations, specially in the low state,
are wanted to consolidate the photometric orbital modulation found in this work, and simultaneous photometric and spectroscopic data would provide means to investigate the 
correlation between the brightness states and the spectral lines behaviour.

\section*{Acknowledgements}

A. S. Oliveira and H. J. F. Lima acknowledge FAPESP -- Funda\c{c}\~{a}o de Amparo \`{a} Pesquisa do Estado de S\~{a}o Paulo -- for financial support under grants 03/12618-7 and 10/12805-5.

\bsp

\label{lastpage}

\end{document}